\documentclass[twocolumn]{aastex701}
\usepackage{xcolor}

\shortauthors{Cordiner et al.}

\accepted{for publication in ApJ Letters, 10th September 2025}

\begin{document}

\title{JWST detection of a carbon dioxide dominated gas coma surrounding interstellar object 3I/ATLAS}

\correspondingauthor{M. A. Cordiner}
\email{martin.cordiner@nasa.gov}

\author[0000-0001-8233-2436]{Martin A. Cordiner}
\affiliation{Astrochemistry Laboratory, NASA Goddard Space Flight Center, 8800 Greenbelt Road, Greenbelt, MD 20771, USA.}
\affiliation{Department of Physics, Catholic University of America, Washington, DC 20064, USA.}
\email{martin.cordiner@nasa.gov}

\author[0000-0002-6006-9574]{Nathan X. Roth}
\affiliation{Astrochemistry Laboratory, NASA Goddard Space Flight Center, 8800 Greenbelt Road, Greenbelt, MD 20771, USA.}
\affiliation{Department of Physics, American University, 4400 Massachusetts Avenue, NW, Washington, DC 20016, USA.}
\email{nathaniel.x.roth@nasa.gov}

\author[0000-0002-6702-7676]{Michael S. P. Kelley}
\affiliation{Department of Astronomy, University of Maryland, College Park, MD 20742-0001, USA.}
\email{msk@astro.umd.edu}

\author[0000-0002-2668-7248]{Dennis Bodewits}
\affiliation{Physics Department, Edmund C. Leach Science Center, Auburn University, Auburn, AL 36849, USA.}
\email{dennis@auburn.edu}

\author[0000-0001-6752-5109]{Steven B. Charnley}
\affiliation{Astrochemistry Laboratory, NASA Goddard Space Flight Center, 8800 Greenbelt Road, Greenbelt, MD 20771, USA.}
\email{steven.b.charnley@nasa.gov}

\author[0000-0001-7479-4948]{Maria N. Drozdovskaya}
\affiliation{Physikalisch-Meteorologisches Observatorium Davos und Weltstrahlungszentrum (PMOD/WRC), Dorfstrasse 33, CH-7260, Davos Dorf, Switzerland.}
\email{maria.drozdovskaya.space@gmail.com}

\author[0000-0003-0774-884X]{Davide Farnocchia}
\affiliation{Jet Propulsion Laboratory, California Institute of Technology, 4800 Oak Grove Dr., Pasadena, CA 91109, USA.}
\email{Davide.Farnocchia@jpl.nasa.gov}

\author[0000-0001-7895-8209]{Marco Micheli}
\affiliation{ESA NEO Coordination Centre, Planetary Defence Office, European Space Agency, Largo Galileo Galilei, 1, 00044 Frascati (RM), Italy.}
\email{marco.bs.it@gmail.com}

\author[0000-0001-7694-4129]{Stefanie N. Milam}
\affiliation{Astrochemistry Laboratory, NASA Goddard Space Flight Center, 8800 Greenbelt Road, Greenbelt, MD 20771, USA.}
\email{stefanie.n.milam@nasa.gov}
\author[0000-0001-7694-4129]{Cyrielle Opitom}
\affiliation{Institute for Astronomy, University of Edinburgh, Royal Observatory, Edinburgh EH9 3HJ, UK.}
\email{copi@roe.ac.uk}

\author[0000-0003-4365-1455]{Megan E. Schwamb}
\affiliation{Astrophysics Research Centre, School of Mathematics and Physics, Queen's University Belfast, Belfast BT7 1NN, UK.}
\email{m.schwamb@qub.ac.uk}

\author[0000-0003-3091-5757]{Cristina A. Thomas}
\affiliation{Northern Arizona University, Department of Astronomy and Planetary Science,
P.O. Box 6010, Flagstaff, AZ, 86011 USA.}
\email{cristina.thomas@nau.edu}

\author[0000-0002-7156-8029]{Stefano Bagnulo}
\affiliation{Armagh Observatory and Planetarium, College Hill, Armagh, BT61 9DG, UK.}
\email{Stefano.Bagnulo@Armagh.ac.uk}

\begin{abstract}

3I/ATLAS is the third confirmed interstellar object to visit our Solar System, and only the second to display a clear coma. Infrared spectroscopy with the James Webb Space Telescope (JWST) provides the opportunity to measure its coma composition and determine the primary activity drivers. We report the first results from our JWST NIRSpec campaign for 3I/ATLAS, at an inbound heliocentric distance of $r_H=3.32$~au. The spectral images (spanning $0.6$--$5.3$~$\mu$m) reveal a CO$_2$ dominated coma, with enhanced outgassing in the sunward direction, and the presence of H$_2$O, CO, water ice, dust {and a tentative detection of OCS}. The coma CO$_2$/H$_2$O mixing ratio of $7.6\pm0.3$ is among the highest ever observed in a comet, and is $4.5\sigma$ above the trend as a function of $r_H$ for long-period and Jupiter-family comets (excluding the outlier C/2016 R2). Our observations are compatible with an intrinsically CO$_2$-rich nucleus, which may indicate that 3I/ATLAS contains ices exposed to higher levels of radiation than Solar System comets, or that it formed close to the CO$_2$ ice line in its parent protoplanetary disk. A low coma H$_2$O gas abundance may also be implied, for example, due to inhibited heat penetration into the nucleus, which could suppress the H$_2$O sublimation rate relative to CO$_2$ and CO.

\end{abstract}


\section{Introduction} 

Comets and planetesimals are theorized to form in large numbers during the accretion of planetary systems. Many of these small bodies, composed of ice, rock, and dust, are subsequently expected to be ejected into interstellar space through gravitational encounters with larger, planetary or stellar, bodies \citep{2004come.book..153Dones,2010Sci...329..187Levison,2018MNRAS.476.3031R,2020ApJ...897...60Pfalzner,2025MNRAS.537.3123Zheng}. The apparitions of the first confirmed interstellar objects (ISOs) 1I/`Oumuamua in October 2017 and 2I/Borisov in August 2019 provided confirmation of this theory, and offered an unprecedented opportunity to study the nature of matter delivered to our Solar System from a distant planetary system, thus spawning a new field of planetary science.

Due to its faintness and short observing window, spectroscopic characterization of volatiles from 1I/`Oumuamua proved elusive, so the composition of this object remains highly uncertain \citep{2019NatAs...3..594O}. On the other hand, for the intrinsically brighter and more active 2I/Borisov, ultraviolet, optical, and submillimeter observations \citep[\emph{e.g.}][]{2020NatAs...4..861Cordiner,2020NatAs...4..867Bodewits,guzik2021gaseous-483,opitom21, Xing2020} provided intriguing glimpses of its coma composition, and revealed an object that was similar in many ways to the well-studied comets from our own Solar System, but with an unusually strong enrichment in carbon monoxide (CO) \citep{2020NatAs...4..861Cordiner,2020NatAs...4..867Bodewits}. Considering the difficulty of studying the ices in the midplanes of protoplanetary disks and planetary systems elsewhere in our Galaxy, continued spectroscopic observations of interstellar objects have the potential to reveal crucial details on the physics and chemistry of planet formation in planetary systems other than our own. 

The discovery of a third interstellar object (3I/ATLAS) was announced on 2025 July 1 by the Asteroid Terrestrial-impact Last Alert System \citep[ATLAS;][]{2018PASP..130f4505Tonry}. Based on its inbound orbital eccentricity ($6.144 \pm 0.016$) and heliocentric radial velocity projected to infinity ($57.95 \pm 0.05$~km~s$^{-1}$) \citep{2025ApJ...989L..36Seligman}, 3I/ATLAS has been confirmed to be on a gravitationally unbound, hyperbolic interstellar trajectory. Dynamical modeling of a population of Galactic interstellar objects (ISOs) shows that the high velocity of 3I/ATLAS is consistent with a relatively large dynamical age of 3--11 billion years \citep{hopkins25,taylor25}. This age, coupled with its trajectory, implies 3I/ATLAS could have originated from a relatively old, low-metallicity stellar system, plausibly from the kinematically hot, ``thick disk'' population of the Milky Way. Chemical differences between the volatile content of 3I/ATLAS and our Solar System's comets may therefore be expected.

Similar to 2I/Borisov, 3I/ATLAS has been shown to display clear cometary activity \citep{2025ATel17263....1Jewitt,2025ATel17264....1Alarcon,2025ATel17275....1Minev}. Early spectroscopic and photometric observations revealed a compact nucleus (effective radius $<2.8$~km), a  bright, dusty coma, with a dust mass-loss rate of 12--120~kg\,s$^{-1}$ \citep{Jewitt2025}, and a red spectral slope, with possible water ice absorption at 2.0~\micron{} \citep{kareta25, 2025arXiv250714916Yang}. The Neil Gehrels-Swift Observatory detected ultraviolet emission from the gas-phase OH radical \citep{2025arXiv250804675Xing}, and assuming H$_2$O to be the photolysis parent, a water production rate of $(1.36 \pm 0.35)\times 10^{27}$ s$^{-1}$ was derived at a heliocentric distance of $r_H = 2.9$ au.

 Here, we present the first set of infrared spectroscopic observations from our campaign to observe 3I/ATLAS using the James Webb Space Telescope (JWST). This article focuses on analyzing the rovibrational fluorescence emission from H$_2$O, CO$_2$, and CO, which are the main drivers of coma activity in typical Solar System comets. Molecular production rates and mixing ratios are derived, enabling chemical characterization of the object's volatile gas inventory.

\section{Observations}

Observations of 3I/ATLAS were performed using JWST \citep{gardner23} on UT 2025-08-06 between 11:02--11:20, using the NIRSpec integral field unit (IFU; \citealt{boker22}), as part of program ID 5094. {The PRISM dispersive element was used, resulting in a $30\times30$ array of spectra} covering $\lambda = 0.6$--5.3 $\mu$m, with a resolving power $R_\lambda=\lambda/\Delta\lambda$ that varies from $\approx30$ at 1.2~$\mu$m to $\approx300$ at 5.3~$\mu$m. {The IFU pixel size is $0\farcs1$, which is approximately the same as the FWHM of the JWST point-spread function at 3~$\mu$m.}

3I/ATLAS was acquired and tracked in the IFU using JPL Horizons ephemeris solution $\#19$, {when the object was 2.73~au from the telescope, at $r_H=3.32$~au and a phase angle of $16.1^{\circ}$}. The total on-target exposure duration was 640~s, divided across four dither positions, each spatially separated (in the approximate shape of a square) with offsets of $\sim0\farcs2$ from the (central) targeted position. The data were reduced using the JWST Calibration Pipeline software version v1.19.1 \citep{bushouse25} using the JWST Calibration Reference Data System context file 1413.  The four dithers were shifted and combined in the rest frame of the comet during image processing, thus allowing detector artifacts and cosmic rays to be identified and removed. A similar set of four exposures of the sky background was obtained, offset by $180''$ along the horizontal axis of the IFU aperture. This allows contamination from background infrared sources, zodiacal light, and the telescope to be identified and subtracted.  Data cubes with and without background subtraction were produced.  The latter allows for the analysis of faint gas emission bands without the contribution of additional noise from a background subtraction.  For each spectral data cube, the pipeline produces uncertainty and data quality maps, which were used during our analysis, in particular, for the derivation of formal ($1\sigma$) error estimates.  After combining the four dithered observations, the absolute calibration accuracy is expected to be 3\%.\footnote{\url{https://jwst-docs.stsci.edu/jwst-calibration-status/nirspec-calibration-status/nirspec-ifu-calibration-status}}

\section{Results}

\begin{figure}
\begin{center}
\includegraphics[width=\columnwidth]{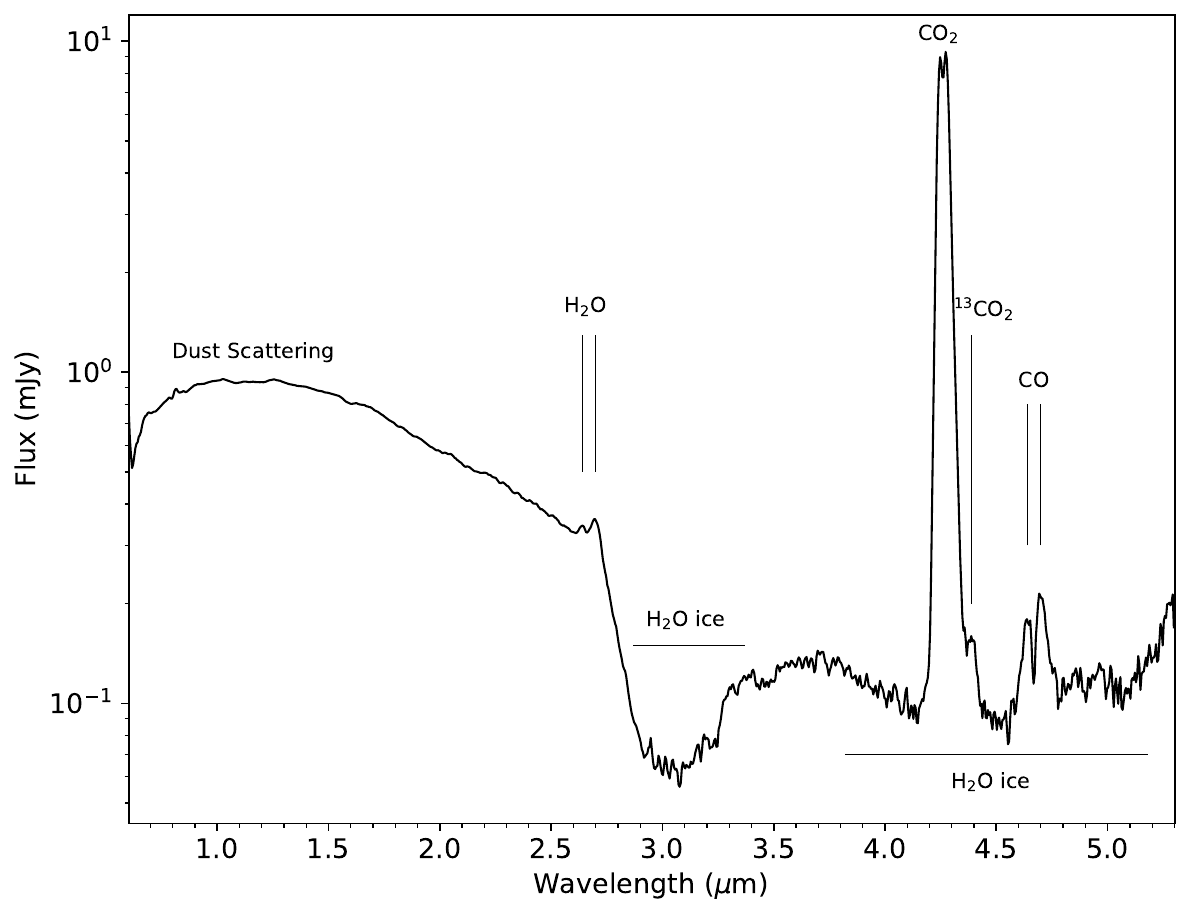}
\end{center}
\caption{JWST spectrum of 3I/ATLAS using the NIRSpec PRISM (sky background subtracted), spatially integrated over the IFU field of view, and plotted with a logarithmic flux scale. Prominent spectral features are labeled.}
\label{fig:spectrum}
\end{figure}

The observed flux was integrated over the entire IFU field of view to produce the spectrum shown in Figure \ref{fig:spectrum}. Prominent features include a broad maximum at around 1.2~$\mu$m, due to scattered sunlight from coma dust grains, and a strong, narrow (double) peak at 4.3~$\mu$m, which is assigned to the main ($\nu_3$) rovibrational emission band of gas-phase CO$_2$. Weaker gas emission bands from H$_2$O ($\nu_1+\nu_3$), CO ($v=1-0$) and $^{13}$CO$_2$ ($\nu_3$) are also present, along with broad absorption features centered around 3.0~$\mu$m and 4.5~\micron, attributed to the OH stretching mode and lattice vibrations, respectively, of H$_2$O ice in the coma --- likely in the form of small ($\lesssim10$ micron-sized) icy grains \citep{moore92,leto03,mastrapa09}.

\begin{figure*}
\begin{center}
\includegraphics[height=6.5cm]{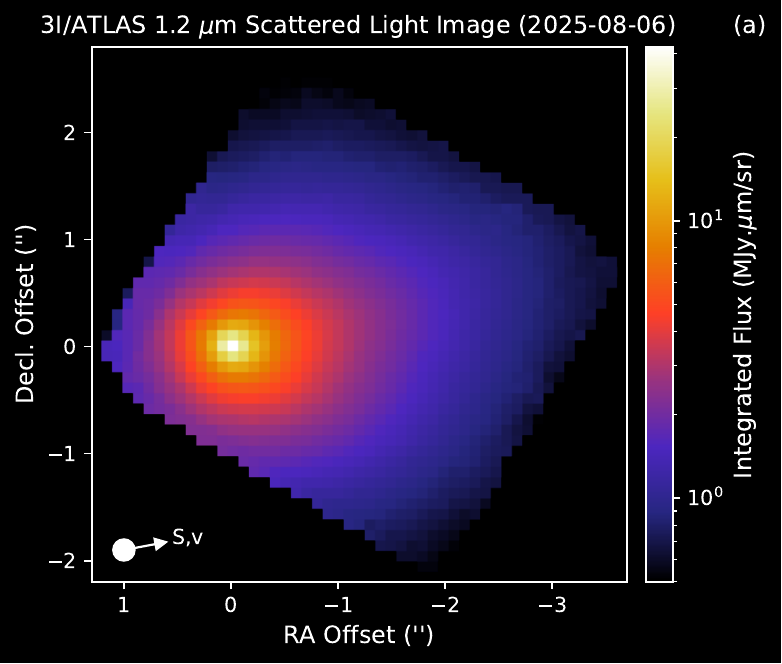}
\includegraphics[height=6.5cm]{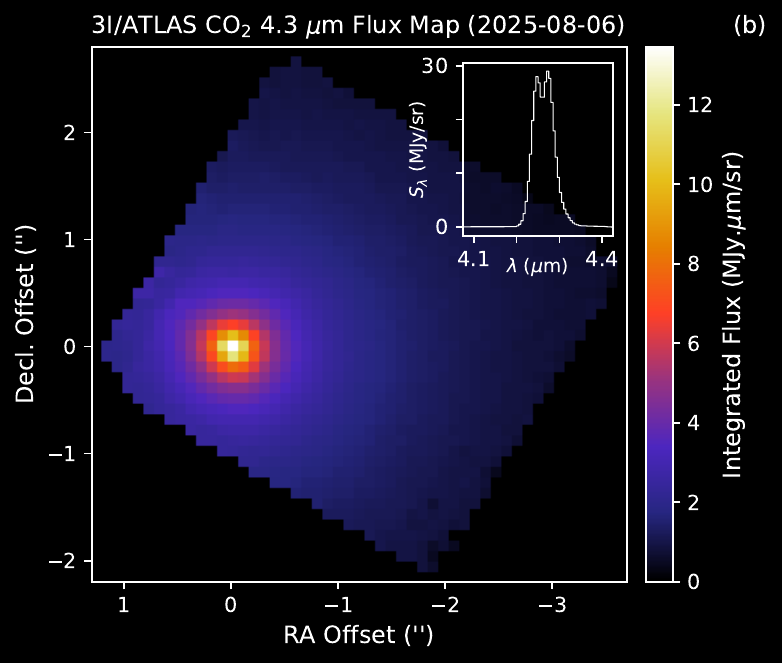}
\includegraphics[height=6.5cm]{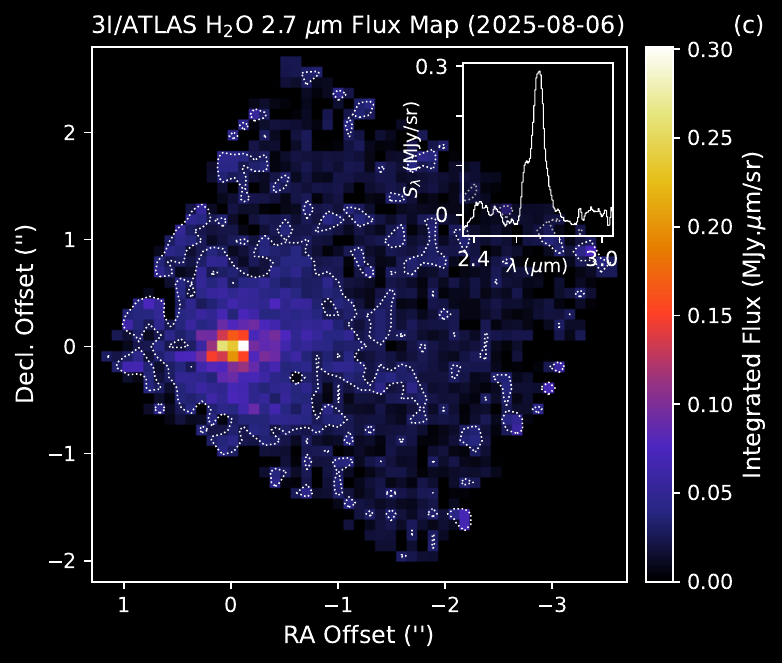}
\includegraphics[height=6.5cm]{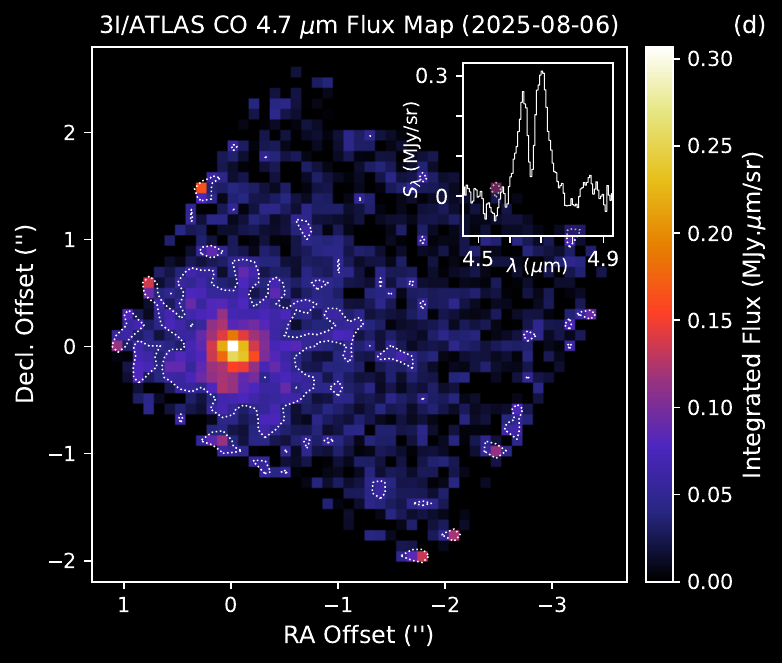}
\end{center}
\caption{Spectrally integrated flux maps for 3I/ATLAS observed using JWST NIRSpec: (a) scattered light from coma dust at $\sim1.2$~$\mu$m, plotted on a logarithmic scale to highlight the coma shape, (b) CO$_2$ at 4.3~$\mu$m, (c) H$_2$O at 2.7~$\mu$m, and (d) CO at 4.7~$\mu$m. {Image axes are aligned with the equatorial (RA/decl.) grid}. Molecular line emission has been isolated by subtracting a polynomial fit to the adjacent continuum. Spatial coordinates are with respect to the brightest pixel in the continuum dust map. For panels (b)--(d), inset plots (upper right) show the continuum-subtracted spectra, spatially averaged across all IFU pixels. Panel (a) lower left corner shows the direction of the (sky-projected) comet-sun (S) and nucleus velocity ($v$) vectors ({which are too close to distinguish}). {For H$_2$O and CO, the $3\sigma$ noise level is shown with a dotted contour; for the dust and CO$_2$ maps, the $3\sigma$ noise level lies outside the IFU boundary, so is not shown.} }
\label{fig:maps}
\end{figure*}

To isolate the CO$_2$, H$_2$O, and CO gas emission features, spectral data were extracted for each IFU pixel within the vicinity of each emission band, and a polynomial continuum fit was performed, excluding the spectral region directly inside each molecular emission band (see Section \ref{sec:modeling}).  For CO$_2$ and CO, a 3rd-order polynomial was used, whereas for H$_2$O a 5th-order polynomial was used, to better fit the wing of the 3.0~$\mu$m ice band .   After continuum subtraction, the IFU spectra were integrated across the detected full emission width of each feature, then plotted as maps in Figure \ref{fig:maps}. The spatially averaged, continuum-subtracted emission band profile for each molecule is shown in the upper-right inset of their respective panels. The $\sim1.2$~$\mu$m scattered light image (integrated between 0.8--1.4~$\mu$m) is also shown.  The continuum brightness at 0.75 and 1.25~\micron{} is 101$\pm$3 and 124$\pm$4 $\mu$Jy inside a $0\farcs8$-diameter circular aperture, corresponding to cometary $Af\rho$ quantities \citep{ahearn84} of 392 and 492~cm, respectively (quoted without a correction for phase darkening).

The gas and dust maps for 3I/ATLAS show a well-defined peak, offset East from the center of the IFU by $1\farcs2$ (Figure \ref{fig:maps}). As shown by a later ephemeris reconstruction, about $0\farcs5$ of the offset can be explained by a $1.7\sigma$ error on the predicted ephemeris, while the remaining $0\farcs7$ offset remains under investigation; {the astrometry available to-date shows no evidence for non-gravitational acceleration of 3I/ATLAS}. The 1.2 $\mu$m scattered light and CO$_2$ emission maps reveal an extended coma of dust and gas that spans the full extent of the NIRSpec IFU.  The H$_2$O and CO emission is weaker, and therefore noisier, but nevertheless confirms the presence of a spatially extended molecular coma. While the scattered light shows a clear asymmetry along the Sun-comet axis --- enhanced in the direction of the sky-projected comet-Sun and velocity vectors (see also \citealt{Jewitt2025, chandler25}) --- the gas distributions (particularly CO$_2$ and CO), appear relatively more symmetrical. 

\begin{figure}
\begin{center}
\includegraphics[width=\columnwidth]{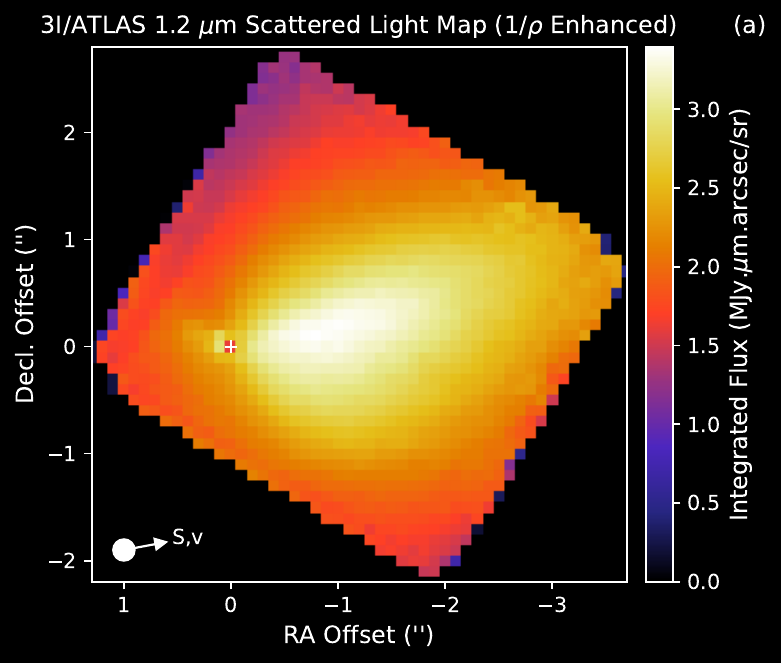}
\end{center}
\caption{$1/\rho$-enhanced $1.2$~$\mu$m scattered light map for 3I/ATLAS observed using JWST NIRSpec. This is panel (a) from Figure \ref{fig:maps}, multiplied by $\rho$ (the sky-projected distance from the center of the brightest pixel --- the assumed location of the nucleus). Similarly enhanced maps of the gas emission are shown in Appendix A (Figure \ref{fig:rhomaps}). {Image axes are aligned with the equatorial (RA/decl.) grid. The white cross shows the position of the nucleus pixel}. }
\label{fig:dust-rhomap}
\end{figure}

To further investigate the coma structure, a $1/\rho$ enhanced version of the dust map (where $\rho$ is the sky-projected nucleocentric distance), is shown in Figure \ref{fig:dust-rhomap}, and similarly for the gas maps in Appendix A (Figure \ref{fig:rhomaps}). In these enhanced maps, the dominant coma spatial feature: $\sim1/\rho$ dilution of the observed column densities due to quasi-spherical expansion, has been divided out. The $1/\rho$-enhanced 1.2~$\mu$m map reveals a strong, plume-like feature emanating from the pseudo-nucleus in the approximate direction of the Sun (slightly north of west), with an additional, weaker feature to the north-east. Since the dust in relatively faint comets such as 3I/ATLAS is optically thin at this wavelength, the shape of this feature is interpreted as resulting from an enhanced coma dust density in the sunward direction, potentially from the fragmentation of dust grains increasing the scattering cross section with distance from the nucleus \citep{combi94,jones08}.  The gas maps on the other hand, show more subtle asymmetries. CO$_2$ exhibits an azimuthal minimum towards the north, that we attribute to a combination of weak coma sub-structure and optical depth effects, since CO$_2$ becomes optically thick close to the nucleus. 

Taking the ratio of continuum-subtracted fluxes within a {$0\farcs4$-radius circular aperture centered $0\farcs6$ from the brightest pixel on the sunward ($S$) and antisunward ($S'$) sides, gives $S/S'=1.54$ for the (1.2~$\mu$m) dust, 1.03 for CO$_2$, 1.31 for H$_2$O and 1.00 for CO} (with uncertainties of $<1\%$ on all measurements). Our observations thus reveal a heterogeneous coma morphology consistent with different outgassing patterns for the different species.  Such heterogeneity can be explained by various factors, such as the different molecular sublimation temperatures, nucleus release and coma acceleration mechanisms, as well as differing inertial/fluid-dynamical properties for the gas and dust. The degree of asymmetry of the observed gases is likely related to their sublimation temperatures ($T_{sub}$), with $S/S'({\rm H_2O})>S/S'({\rm CO_2})>S/S'({\rm CO})$, congruent with  $T_{sub}({\rm H_2O})>T_{sub}({\rm CO_2})>T_{sub}({\rm CO})$ \citep{womack17}. This implies that the sublimation of CO$_2$ and CO is more fully activated than H$_2$O.  Full interpretation of the observed outgassing morphologies, including the surprisingly strong sunward dust enhancement, will require detailed physical modeling. Nevertheless, our data are consistent with the origin of this dust feature being influenced by enhanced gas sublimation (and therefore, outgassing) rates associated with the higher temperature on the dayside of the nucleus. 

 The 3.0~\micron{} H$_2$O ice band depth is 78.3$\pm$0.2\% at 2.9--3.1~\micron{}, with respect to the continuum at 2.5 and 3.8~\micron.  The band remains strong across the field of view, varying by no more than 10\%. Water ice also has broad near-infrared absorption features at 1.5, 2.0, 3.0, and 4.5~\micron. The 4.5~\micron{} band is difficult to measure in our data due to blending with CO$_2$ and CO gas emission and likely thermal continuum, the 2.0~\micron{} band appears very weak, and there is no discernible 1.5~\micron{} band. Furthermore, the 2.0~\micron{} spectral region in our combined IFU dataset is affected by several bad pixels. {The average 2.0~\micron{} band depth in the two dither positions with clean spectra in this region is (1.8$\pm$0.1)\%. This was estimated by normalizing the reflectance spectrum within a 0\farcs4 radius aperture with a linear fit between 1.75 and 2.22~\micron{}, and measuring the mean value at 1.95--2.05~\micron.  The appearance of a strong 3-\micron{} band, but weak or absent 1.5- and 2.0-\micron{} bands implies the ice grains are micrometer size or smaller, while the shape of the band may be consistent with contributions from crystalline as well as amorphous ice \citep{protopapa14}.}
 
 
 


\section{Spectral Modeling}
\label{sec:modeling}

\begin{figure*}
\begin{center}
\includegraphics[width=0.32\textwidth]{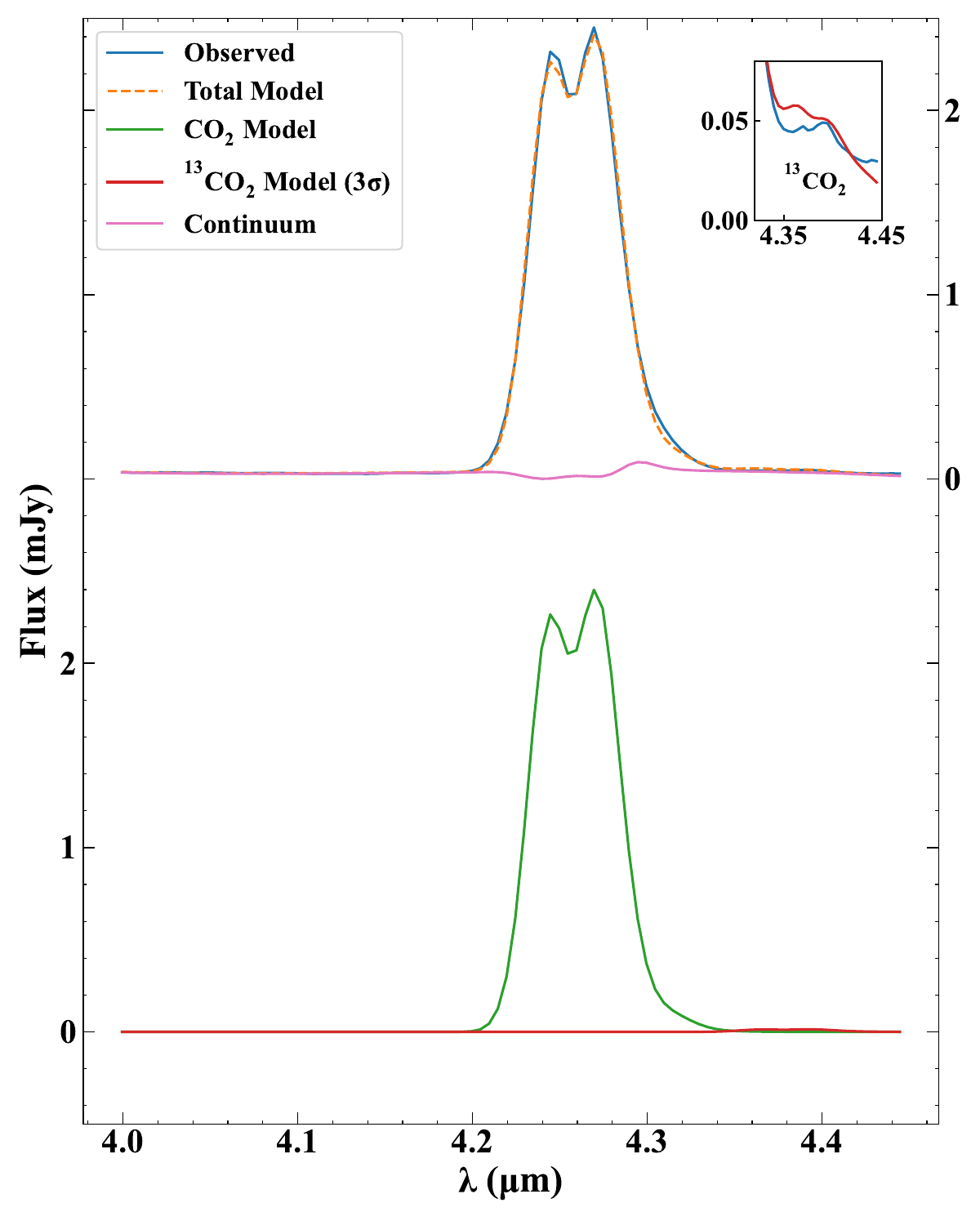}
\includegraphics[width=0.32\textwidth]{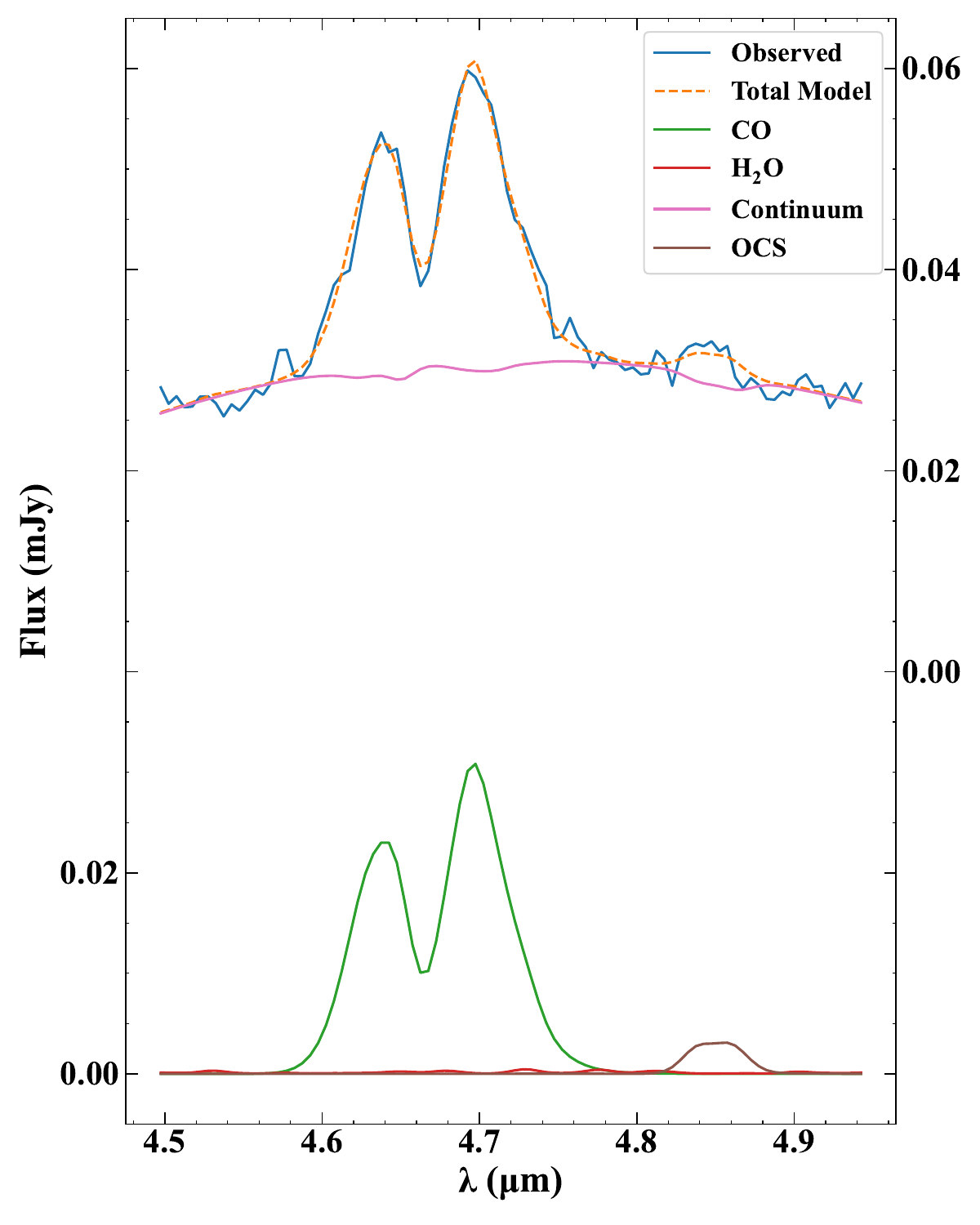}
\includegraphics[width=0.32\textwidth]{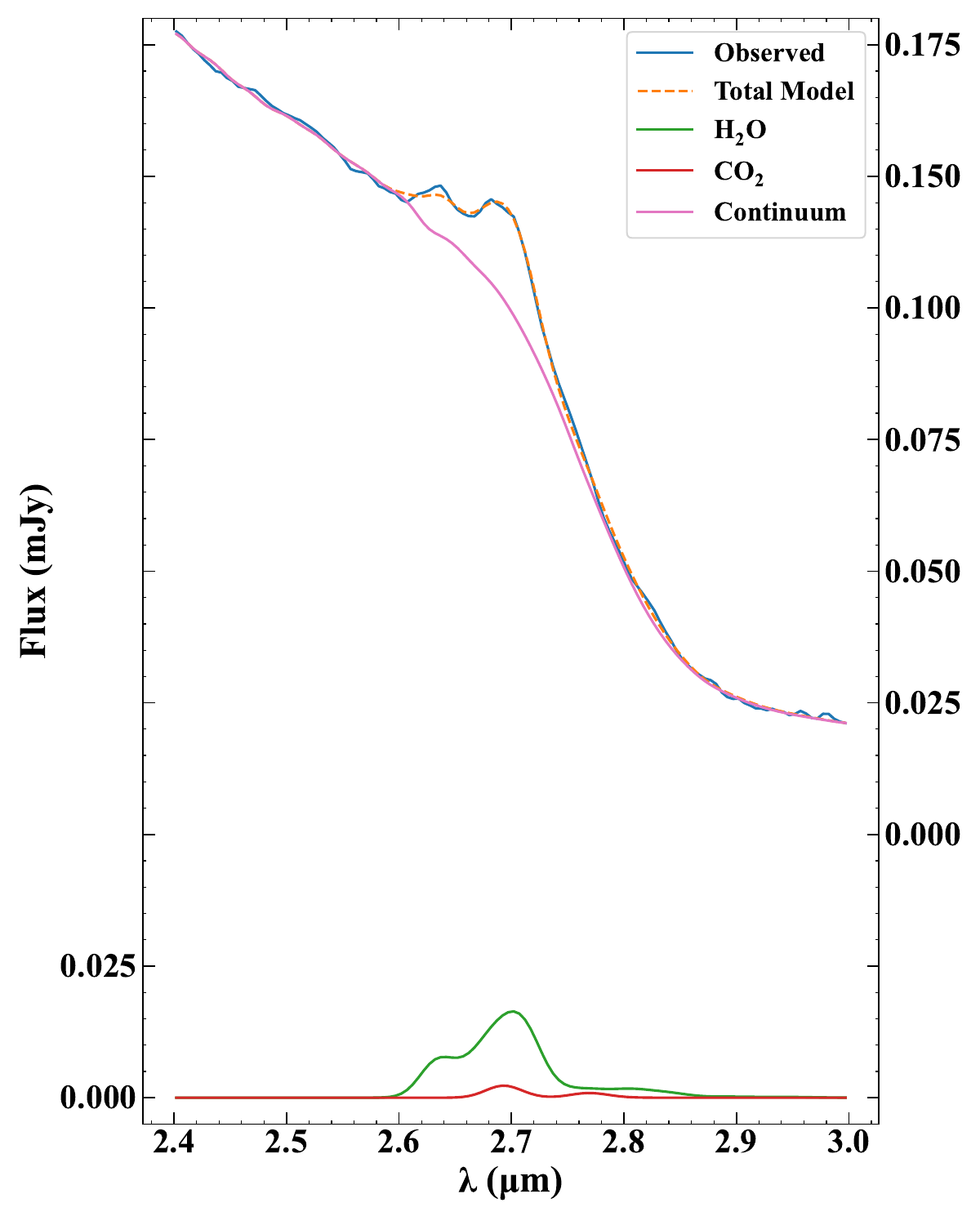}
\end{center}
\caption{Observed NIRSpec molecular spectra of 3I/ATLAS extracted within a $0\farcs625$-radius circular aperture centered on the (pseudo-)nucleus, along with best-fitting gas and continuum models. {The left ordinate axes apply to the model gas component fluxes (lower traces), whereas the right ordinates are for the observed spectrum, total model and continuum model components (upper traces). The weak $^{13}$CO$_2$ band at 4.4~$\mu$m is shown in the inset panel next to the main CO$_2$ band, with $3\sigma$ upper limit spectral model overlaid}.}
\label{fig:fits}
\end{figure*}

To derive gas production rates ($Q$) and rotational temperatures ($T_{rot}$), the background-subtracted IFU data for CO$_2$, CO, and H$_2$O were subject to spectral modeling using optimal estimation routines as part of the Planetary Spectrum Generator (PSG; \citealt{villanueva18}). Initially, we constructed spectral models by taking the average spectrum for each gas inside a $\rho=0\farcs625$ (1240~km) circular aperture centered on the nucleus. The relatively high signal-to-noise ratio (SNR) in this IFU region assisted in helping define the choice of continuum shape and the gases to be included in the model. Figure \ref{fig:fits} shows the integrated spectra within this aperture, along with the best-fitting PSG model gas emission components. For the $4.7$~$\mu$m region, we also {tentatively} identified an emission band at 4.85~$\mu$m consistent with OCS, which was included in the fit. Additional details on the continuum fitting and spectral modeling procedure are given in Appendix B.

\begin{figure*}
\begin{center}
\includegraphics[width=0.7\textwidth]{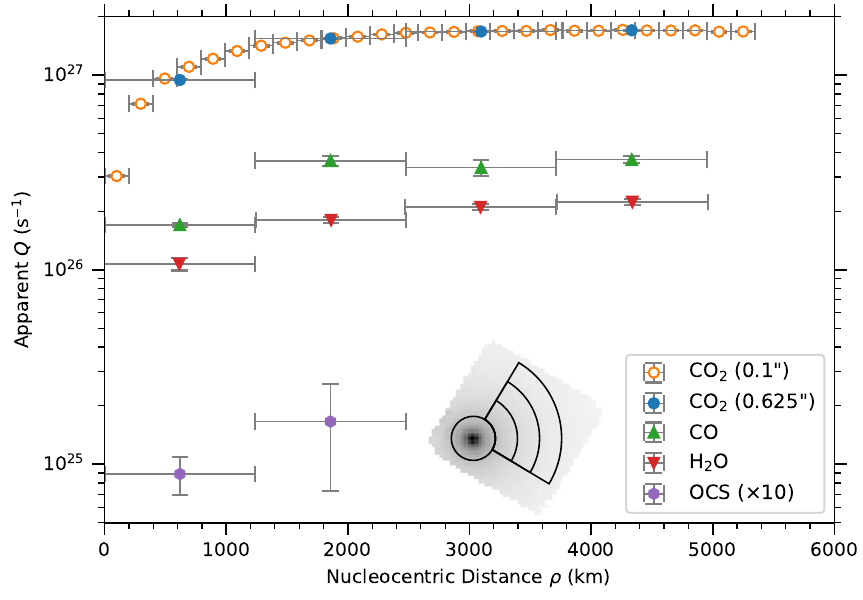}
\end{center}
\caption{Best fitting production rates for CO$_2$, CO, H$_2$O and OCS as a function of distance from the nucleus ($Q$ curves). Spectra were extracted and modeled within four spatial regions of the IFU, as shown in the inset diagram, consisting of a central $0\farcs625$-radius circle, followed by three successive partial annular sectors, each with a $0\farcs625$ radial extent. Due to its higher SNR, CO$_2$ was also modeled within successive $0\farcs1$ annuli surrounding the nucleus. OCS values have been scaled up by a factor of 10 for display. Vertical error bars indicate $1\sigma$ statistical uncertainties, while the horizontal error bars indicate the radial extent of each spatial region.}
\label{fig:qcurves}
\end{figure*}

After modeling the nucleus-centered IFU extract, we proceeded to derive production rates and rotational temperatures as a function of $\rho$, by extracting and modeling the data within successive $0\farcs625$-wide annular sectors, as shown in the inset diagram of Figure~\ref{fig:qcurves}. The resulting production rates as a function of $\rho$ (referred to as ``$Q$ curves'') are also shown in Figure~\ref{fig:qcurves}

For the nucleus-centered extract ($1\farcs25$-diameter circle), our best-fitting models give $Q({\rm CO_2})=(9.50\pm0.05)\times10^{26}$~s$^{-1}$, $Q({\rm CO})=(1.70\pm0.04)\times10^{26}$~s$^{-1}$ and $Q({\rm H_2O})=(1.07\pm0.08)\times10^{26}$~s$^{-1}$ , and $Q({\rm OCS})=(8.9\pm2.0)\times10^{23}$~s$^{-1}$. From the $Q$-curve analysis, ``terminal'' gas production rates were derived for the annular sectors furthest from the nucleus. These provide an improved view of the coma mixing ratios, avoiding the optical depth effects that impact CO$_2$ in the nucleus-centered extract, although they could contain additional contributions from extended/distributed coma sources. The resulting (whole-coma equivalent) terminal gas production rates are $Q({\rm CO_2})=(1.70\pm0.01)\times10^{27}$~s$^{-1}$, $Q({\rm CO})=(3.7\pm0.2)\times10^{26}$~s$^{-1}$, $Q({\rm H_2O})=(2.23\pm0.08)\times10^{26}$~s$^{-1}$, and $Q({\rm OCS})=(1.7\pm0.9)\times10^{24}$~s$^{-1}$.

The $Q$ curves for CO$_2$ and CO (Figure \ref{fig:qcurves}) level off towards larger nucleocentric distances, indicating that gas production for these species is confined within $\sim3000$~km of the nucleus. On the other hand, the H$_2$O $Q$-curve shows no clear asymptote, and the error bars allow for the possibility of continued H$_2$O production towards the edge of the NIRSpec IFU. Therefore, we cannot rule out a contribution to H$_2$O gas in the outer coma from sublimating icy grains, which may be expected based on our detection of coma H$_2$O ice. {Our terminal $Q({\rm H_2O})$ value is significantly smaller than the value of $(1.36\pm0.35)\times10^{27}$~s$^{-1}$ measured at $r_H=2.9$~au by \citet{2025arXiv250804675Xing}. This could be due to a rapid increase in H$_2$O production between $r_H=3.3$--2.9~au, with a possible contribution from icy-grain sublimation within the relatively large, 20,000~km diameter Swift aperture.}


Although $^{13}$CO$_2$ is securely detected in the nucleus-centered extract (Figure \ref{fig:fits}), optical depth effects and blending with the $^{12}$CO$_2$ wing preclude the derivation of a reliable $^{13}$CO$_2$ production rate in this region. We attempted to retrieve $Q$($^{13}$CO$_2$) in the annular sector between $0\farcs625$--$1\farcs25$ from the nucleus, but only an upper limit $Q({\rm ^{13}CO_2})<1.8\times10^{25}$~s$^{-1}$ could be obtained, resulting in a ($3\sigma$) lower limit on the $^{12}$C/$^{13}$C ratio of $>63$, which is formally consistent with the terrestrial value of 89 \citep{coplen02}. See Appendix B for further details.

\section{Discussion}

Our JWST NIRSpec observations reveal that 3I/ATLAS contains a substantial volatile ice inventory, with a gas coma unusually rich in CO$_2$ relative to H$_2$O and CO. In the absence of clear detections of other gases, it is reasonable to infer that CO$_2$ outgassing provides the dominant driving force for 3I's nucleus activity, and is responsible for launching dust grains away from the nucleus to produce the distinctive scattered light coma observed at $\sim1.2$~$\mu$m and shorter wavelengths.

\begin{figure*}
\begin{center}
\includegraphics[width=0.7\textwidth]{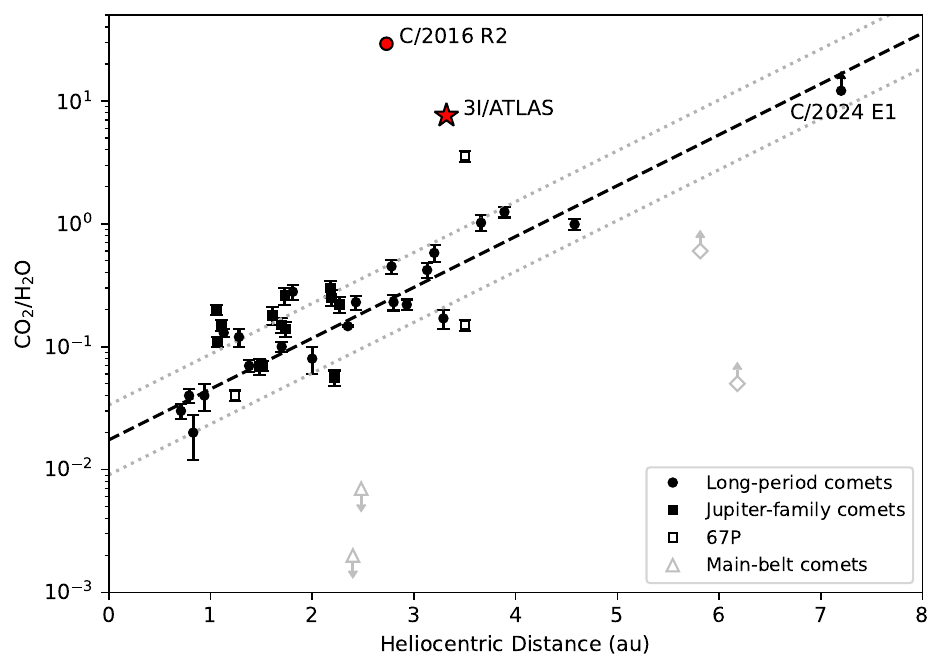}
\end{center}
\caption{Coma CO$_2$/H$_2$O mixing ratios as a function of heliocentric distance for previously-observed comets, grouped by category: (1) long-period comets (LPCs), including Oort Cloud and Halley-type comets, (2) Jupiter-family comets (JFCs), (3) Main-belt comets (MBCs), and (4) Centaurs. Data are from the compilation of \citet{pinto22}, with additional values from \citet{kelley23,pinto23,hsieh25,Woodward2025,snodgrass25}. {Values for 67P/Churyumov–Gerasimenko at $r_H=3.5,1.2,3.5$~au (pre- to post-perihelion) are included, from \citet{combi20}}. Upper and lower limits are shown with arrows. A log-linear curve is fitted to the combined LPC + JFC dataset (dashed line), with $\pm1\sigma$ margins shown as dotted grey lines (where $\sigma$ is the standard deviation of the data from the fit). The peculiar, hypervolatile-rich comet C/2016 R2 was excluded from the fit. 3I/ATLAS (red star) is labeled, in addition to the lower limit for the recently-observed distant Oort cloud comet C/2024 E1 \citep{snodgrass25}.}
\label{fig:co2h2o}
\end{figure*}

For previously-observed comets in our Solar System, the relative coma abundances of CO$_2$, CO, and H$_2$O are known to vary widely (up to several orders of magnitude) between different comets, with some of the variability attributed to the differing relative volatilities of these species as a function of temperature \citep{ootsubo12, AHearn2012,pinto22}. Indeed, variation in the coma CO$_2$/H$_2$O and CO/H$_2$O mixing ratios as a function of heliocentric distance is both theoretically predicted and observed \citep{marboeuf14,pinto22}.  However, compared with previous comets observed at similar heliocentric distances ($r_H\sim3$--4~au), the CO$_2$-dominated outgassing in 3I/ATLAS appears unusual. In Figure \ref{fig:co2h2o}, we plot the CO$_2$/H$_2$O coma mixing ratio measured in previous comets as a function of $r_H$, and draw a log-linear trend line, fitted to the combined dataset of long- and short-period comets (excluding the peculiar outlier C/2016 R2 and the lower limit for C/2024 E1).  The fit is weighted by the data uncertainties; where no uncertainties were available, an error bar of 10\% was assumed. 

The interstellar object 3I/ATLAS has a coma CO$_2$/H$_2$O ratio of $7.6\pm0.3$, which is 18 times larger than expected for its heliocentric distance, based on the fit to previously observed cometary data (Figure \ref{fig:co2h2o}). This corresponds to $4.5\sigma$ away from the trend line, and shows that 3I's coma CO$_2$/H$_2$O ratio is unusually high. The only other comet known to have a CO$_2$/H$_2$O ratio so far outside the normal Solar System trend is C/2016 R2 (PanSTARRS) \citep{mckay2019}. C/2016 R2 is considered to be one of the most peculiar comets ever observed, as a result of its large hypervolatile content \citep{biver18,cordiner22} and correspondingly high CO/H$_2$O ratio.  The CO/H$_2$O ratio of $1.65\pm0.09$ in 3I/ATLAS, on the other hand, is more compatible with previous cometary observations, which have values $<7$ between $r_H=3$--4 au \citep{pinto22}. Intriguingly, our CO/H$_2$O ratio is within the range of values (1.3--1.6) measured in 2I/Borisov at $r_H=2.0$~au, although the H$_2$O production rate was observed to be falling rapidly around the time of those observations \citep{2020NatAs...4..867Bodewits}.

Due to the presence of water ice in cometary comae at $\gtrsim$3~au \citep[\emph{e.g.}][]{lellouch98,kawakita04,protopapa18}, previous measurements of the gas CO$_2$/H$_2$O ratio using larger spectroscopic apertures \citep[\emph{e.g.}][]{ootsubo12} could have been impacted by icy grain sublimation, thus reducing the observed ratio. It is difficult to assess the full impact this would have on the points around $r_H=3$--4 au in Figure \ref{fig:co2h2o}, {so further observations of 3I/ATLAS are recommended, closer to perihelion.}

{The active sublimating surface area for each of our detected gases is calculated in Appendix C.} Following the analysis of \citet{Jewitt2025}, the CO$_2$ active area of 3.1~km$^2$ is sufficient to drive the development of 3I's observed dust coma, even if the dust grains are relatively large ($\sim100$~$\mu$m) in size. The relatively small active area for H$_2$O (2.2~km$^2$) could be partly explained by the relatively high sublimation temperature of H$_2$O \citep{womack17}, if the internal temperature of most of the nucleus ($T_{nuc}$) was in the range $T_{sub}({\rm H_2O})> T_{nuc} > T_{sub}({\rm CO_2})$ (but still close enough to $T_{sub}({\rm H_2O})$ to allow some H$_2$O sublimation), at the time of our observations. Sublimation of H$_2$O may therefore become more fully activated as 3I/ATLAS moves closer to the Sun \citep{puzia25}, in which case a more accurate picture of the nucleus composition will be obtained. The very high CO$_2$/H$_2$O ratio observed by JWST could therefore indicate that $T_{nuc}$ is lower than that experienced by typical Solar System comets at a similar $r_H$. This could arise as a result of a higher albedo or lower thermal conductivity of the nucleus surface layer compared with typical comets, leading to reduced heating or heat penetration. Higher albedo could be caused by a more ice-rich surface composition than normal, whereas lower thermal conductivity could arise from the presence of a volatile-depleted crust/mantle \citep{guilbert15}. The latter was hypothesized for 1I/`Oumuamua, as a result of irradiation by cosmic rays during the object's interstellar passage \citep{fitzsimmons18}. Accounting for the lower volatility of H$_2$O using the \citep{Cowan1979} ice sublimation model, we predict CO$_2$/H$_2$O $\sim 3.2$ at $r_H=1$~au, which is still an order of magnitude larger than other comets observed near $r_H=1$~au (Figure \ref{fig:co2h2o}).

CO$_2$/H$_2$O ratios greater than unity have been only rarely observed in previous comets. This is likely due to a combination of factors, including (1) relatively sparse number statistics due to the difficulty of observing comets beyond $r_H\gtrsim3$~au (where H$_2$O sublimation is strongly suppressed), and (2) the difficulty of CO$_2$ observations in the pre-JWST era, due to telluric obscuration in the 4.3~$\mu$m region. Furthermore, bulk cometary CO$_2$/H$_2$O ice abundances are typically less than a few tens of percent \citep{boogert15}; a median coma CO$_2$/H$_2$O ratio of 17\% was measured by \citet{ootsubo12}, and a bulk, mission-integrated value of 7\% was derived for comet 67P/Churyumov–Gerasimenko \citep{lauter20}. Indeed, considering the theory that a significant fraction of cometary ice originates in the interstellar medium \citep{ehren00,2016MNRAS.462..977Drozdovskaya}, where CO$_2$/H$_2$O ratios are $\sim10$--50 \% \citep{boogert15,smith25}, a bulk CO$_2$/H$_2$O ratio in excess of unity for 3I/ATLAS would be surprising, perhaps hinting at an unusual, carbon-rich chemical composition for this object.


CO$_2$ is thought to form during the interstellar and protoplanetary disk phases of star formation as a result of reactions between CO and OH on dust grain surfaces. The formation of CO$_2$ competes with the formation of H$_2$O from OH + H reactions \citep{2011ApJ...735..121Noble}. Under dark, non-irradiated conditions, the barrierless hydrogenation reaction resulting in H$_2$O dominates. However, upon exposure to UV radiation and cosmic rays, physicochemical models show that CO$_2$ may start to dominate the ice inventory \citep{2016MNRAS.462..977Drozdovskaya}. \citet{2021A&A...650A.180Notsu} determined that CO$_2$ ice abundances are maximized at moderate ($10^{29}-10^{30}$ erg s$^{-1}$) X-ray luminosities in protostellar envelopes.  Various combinations of physicochemical evolutionary scenarios can theoretically produce zones with CO$_2$/H$_2$O $>1$ in protoplanetary disks, for example: (1) beyond 30 au in the midplanes of larger, UV-irradiated, infall-dominated disks \citep{2016MNRAS.462..977Drozdovskaya}, (2) at a few au over longer timescales due to cosmic ray effects \citep{2018A&A...613A..14Eistrup}, or (3) in elevated disk layers due to UV irradiation from the central protostar \citep{furuya22}. Furthermore, as shown by \citet{stevenson88}, diffusion of sublimated gas outward across the ice line, where it subsequently freezes out, can result in significantly enhanced abundances of that ice.  \citet{price21} modeled this effect for CO in the presence of inward-drifting icy pebbles, to explain the high CO/H$_2$O ratio 2I/Borisov, so a similar enrichment of solid CO$_2$ may be expected in protoplanetary disks, just outside the CO$_2$ ice line. An intrinsically CO$_2$-rich composition for a fraction of the interstellar object population, formed in such regions, may therefore result. Additional theoretical modeling will be required to determine whether a high CO$_2$/H$_2$O ratio could be compatible with an origin for 3I/ATLAS in the low metallicity, thick-disk Galactic stellar population, as suggested by \citet{hopkins25}.

\section{Conclusion}

We performed JWST NIRSpec imaging spectroscopy of the interstellar object 3I/ATLAS at $r_H=3.32$~au on the inbound leg of its flight through the Solar System. Rovibrational emission bands were detected of CO$_2$, CO and H$_2$O, in addition to dust and ice solid-state features, demonstrating the presence of a substantial, gas- and ice-rich coma comparable to that of comets from our own Solar System. The CO$_2$ band at 4.3~$\mu$m was particularly strong. The CO$_2$/H$_2$O mixing ratio of $7.6\pm0.3$ is $4.5\sigma$ above the trend as a function of $r_H$ observed in long-period and Jupiter-family comets (excluding the peculiar C/2016 R2), and suggests the possibility of an intrinsically CO$_2$-rich nucleus. Such a high CO$_2$/H$_2$O ratio has never before been observed in a comet between $r_H=3$--4 au. The combined capabilities of the JWST and Vera C. Rubin Legacy Survey of Space and Time \citep{ivezic19} will facilitate additional observations of Solar System comets at such distances, to help improve the statistics and confirm whether 3I/ATLAS is as unusual as it appears. 

A low coma H$_2$O abundance could also be implied by our data, possibly arising as a result of reduced heat penetration through an unusually thick, insulating crust or mantle. In that case, the sublimation of the less volatile H$_2$O ice could be inhibited relative to the (more volatile) CO$_2$ and CO ices. Further observations at distances $r_H<3$ au will be needed, to facilitate measurement of the bulk nucleus composition of 3I/ATLAS as it passes closer to the Sun and the sublimation of H$_2$O (and other low-volatility ices) becomes more fully activated.

\software{George \citep{ambikasaran15}, JWST Calibration Pipeline software version v1.19.1 \citep{bushouse25}, Matplotlib \citep{hunter2007}, Numpy \citep{harris2020},
Small-Bodies-Node/ice-sublimation \citep{VanSelous2021}, Planetary Spectrum Generator (PSG; \citealt{villanueva18}), sbpy \citep{mommert19}}

\begin{acknowledgments}
This work is based on observations made with the NASA/ESA/CSA James Webb Space Telescope. The data were obtained from the Mikulski Archive for Space Telescopes at the Space Telescope Science Institute, which is operated by the Association of Universities for Research in Astronomy, Inc., under NASA contract NAS 5-03127 for JWST. Astrometric measurements of 3I/ATLAS were obtained using the ESO VLT FORS2 instrument at the La Silla Paranal Observatory, under program ID 115.29F8. We gratefully acknowledge the assistance of other optical observers who submitted astrometric observations of 3I/ATLAS in the weeks leading up to our observations, to help refine the ephemeris position --- in particular, T. Lister, M. Banister, Q. Ye, and D. Seligman. John Stansberry is acknowledged for his assistance with scheduling the JWST observations. Supporting astrometric observations were obtained by the Comet Chasers school outreach program (\url{https://www.cometchasers. org/}), led by Helen Usher, which is funded by the UK Science and Technology Facilities Council (via the DeepSpace2DeepImpact Project), the Open University and Cardiff University. It accesses the LCOGT telescopes through the Schools Observatory/Faulkes Telescope Project (TSO2025A-00 DFET-The Schools' Observatory), which is partly funded by the Dill Faulkes Educational Trust, and through the LCO Global Sky Partners Programme (LCOEPO2023B-013). Observers included E. Maciulis, A. Bankole, J. Bowker,  A. Trawiki, K. Golabek, L. Garrett, O. Roberts, T. Oladunjoye, participants on the British Astronomical Associations' Work Experience project 2025, and representatives from the following schools and clubs: The Coopers Company \& Coborn School; Upminster, UK; Ysgol Gyfun Gymraeg Bro Edern, Cardiff, UK; St Marys Catholic Primary School, Bridgend, UK; Institut d'Alcarràs, Catalonia, Spain; Louis Cruis Astronomy Club, Brazil; Srednja škola Jelkovec (Jelkovec High School), Zagreb, Croatia. D.F. conducted this research at the Jet Propulsion Laboratory, California Institute of Technology, under a contract with the National Aeronautics and Space Administration (80NM0018D0004).  This research has made use of NASA’s Astrophysics Data System Bibliographic Services. This research has made use of data and/or services provided by the International Astronomical Union's Minor Planet Center. M.E.S. acknowledge support in part from UK Science and Technology Facilities Council (STFC) grant ST/X001253/1.

Data Access Statement: All JWST data are available through the Mikulski Archive for Space Telescopes at the Space Telescope Science Institute under proposal ID \#5094 (https://doi.org/10.17909/1jvn-1z72). The data products are under a three month embargo. 
\end{acknowledgments}

\bibliography{refs}{}
\bibliographystyle{aasjournalv7}

\appendix
\section{Radially-Enhanced Flux Maps}
\begin{figure*}
\begin{center}
\includegraphics[height=6.5cm]{Reflected_1.2um_rhoMap.pdf}
\includegraphics[height=6.5cm]{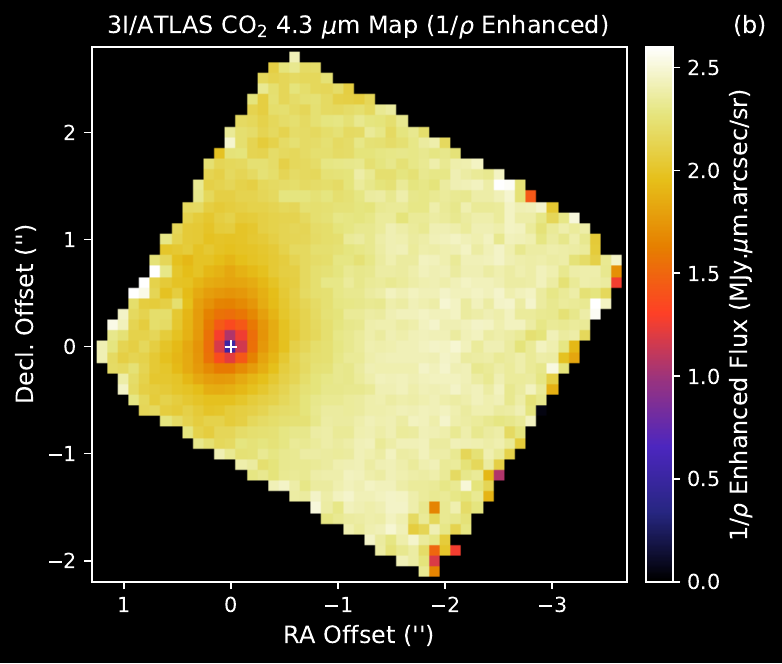}
\includegraphics[height=6.5cm]{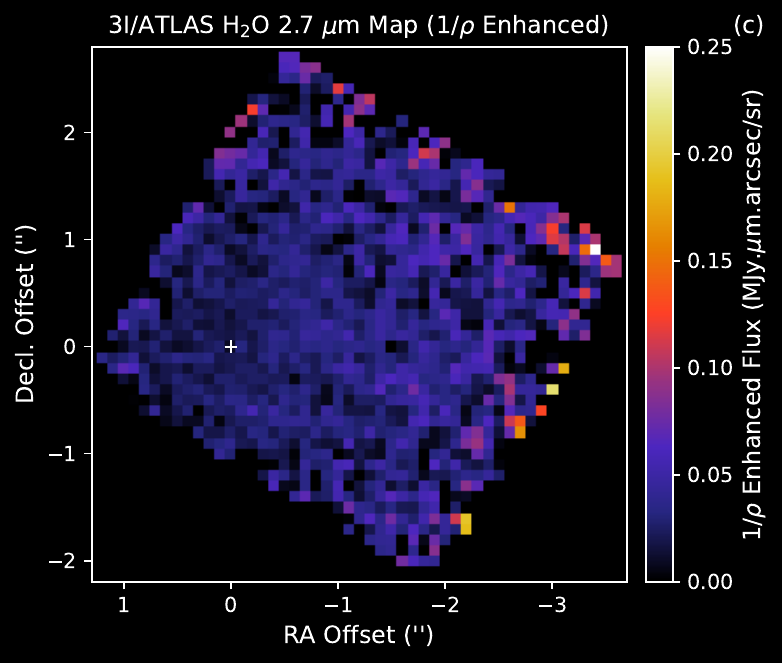}
\includegraphics[height=6.5cm]{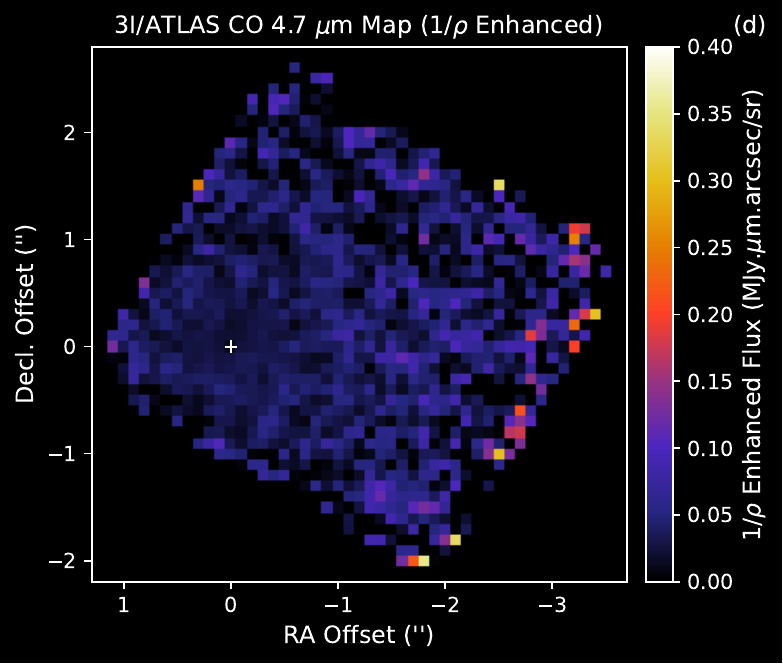}
\end{center}
\caption{$1/\rho$-enhanced flux maps for 3I/ATLAS observed using JWST NIRSpec. These are the images from Figure \ref{fig:maps}, multiplied by $\rho$ (the sky-projected distance from the center of the brightest pixel), for (a) scattered light at $1.2$~$\mu$m, (b) CO$_2$ at 4.3~$\mu$m, (c) H$_2$O at 2.7~$\mu$m, and (d) CO at 4.7~$\mu$m. {Image axes are aligned with the equatorial (RA/decl.) grid}. Panel (a) lower left corner shows the direction of the (sky-projected) comet-sun (S) and nucleus velocity ($v$) vectors ({which are too close to distinguish}). The white cross shows the position of the nucleus pixel.}
\label{fig:rhomaps}
\end{figure*}

Figure \ref{fig:rhomaps} shows ``$1/\rho$ enhanced'' versions of the flux maps in Figure \ref{fig:maps}, where the dilution of the observed column density due to quasi-spherical expansion of the coma (which is proportional to $1/\rho$, where $\rho$ is the sky-projected distance from the center of the nucleus pixel), has been divided out.  The average value of $\rho$ was calculated within each pixel using a $10\times10$-point cartesian sub-sampling.

\section{Spectral Modeling}

Gas production rates ($Q$) and rotational temperatures ($T_{rot}$), were derived as a function of distance from the nucleus for CO$_2$, CO, and H$_2$O, using optimal estimation routines as part of the Planetary Spectrum Generator (PSG; \citealt{villanueva18}, based on synthetic fluorescence models described by \citealt{Villanueva2025}). {We performed baseline fitting to subtract the (nucleus + dust) continuum based on fits to the spectral regions immediately adjacent to the lines of interest (Figure \ref{fig:fits}), adopting the conservative strategy of obtaining an good fit within the noise, using a minimum number of free parameters.}  A 3rd-order polynomial was found to be sufficient for describing the continua underneath and surrounding the CO$_2$ and CO features, whereas for H$_2$O, we tried various analytic functions to produce a good fit, including the wing of the 3~$\mu$m ice band.  Although a 5th order polynomial was found to be sufficient to reproduce the shape of the continuum in this region, a lack of formal constraints across the 2.6--2.8~$\mu$m region spanning the H$_2$O 2.7~$\mu$m band led to increased uncertainties on $Q({\rm H_2O})$.  Therefore, for H$_2$O, we adopted a more physically constrained continuum model, formed as the product of a linear slope and a sigmoid function:
\begin{equation}
    R(\lambda) = \left( m (\lambda - \lambda_c) + b \right) \left(1 - \frac{L}{1 + e^{-\tau (\lambda - \lambda_b)}} \right),
    \label{eq:sigmoid}
\end{equation}
where $\lambda_c=2.5~\micron$ is the linear continuum normalization point, $m$ is the linear slope, $b$ is the y-intercept point, $L$ is the depth of the 3-\micron{} band, $\tau$ controls the slope of the band edge, and $\lambda_b$ controls the wavelength of the band edge. The (variable) exponential onset of the sigmoid function effectively matches the quasi-Gaussian shape of the blue wing of the 3~$\mu$m H$_2$O ice band \citep{leto03}. {By simultaneously optimizing the continuum and spectral line models, uncertainties in the fitted continuum shapes were included in the uncertainties derived for our best-fitting production rates. After experimenting with alternative functional forms for the continuum in the vicinity of our observed spectral lines, we found that the derived production rates remained consistent, within the errors.}

The CO$_2$ and CO spectral regions were modeled using the methods described by \cite{Woodward2025}. The gas outflow velocity was set at 0.44~km\,s$^{-1}$, based on the standard relationship $v = 0.8r_H^{-0.5}$ \citep[\emph{e.g.}][]{ootsubo12}. Molecular photolysis rates appropriate for the active Sun were incorporated from \citet{hue15}. Pixels close to the nucleus can be affected by significant line opacity and PSF-related flux losses, which are difficult to accurately model. Therefore, after modeling the average spectrum within the $\rho=0\farcs625$ nucleus-centered aperture, we proceeded with a ``$Q$-curve'' analysis, deriving the production rates and rotational temperatures as a function of $\rho$ within successive (independent) partial annular sectors centered on the brightest (nucleus-containing) pixel. To avoid flux losses from pixels at the very edge of the IFU, and to focus on a uniform coma angular region, a $90^{\circ}$ inscribed angle was used for all partial annuli, with radial bounds parallel to the north-east and south-east edges of the IFU; see inset diagram in Figure \ref{fig:qcurves}. We found a $0\farcs625$ annulus width to provide sufficient signal-to-noise for all species. Given the very high SNR for CO$_2$, we also generated a higher-resolution $Q$ curve for this species using a $0\farcs1$ annulus width. The resulting production rates and rotational temperatures as a function of $\rho$ are shown in Figure~\ref{fig:qcurves}.

\begin{figure*}
\begin{center}
\includegraphics[width=0.6\textwidth]{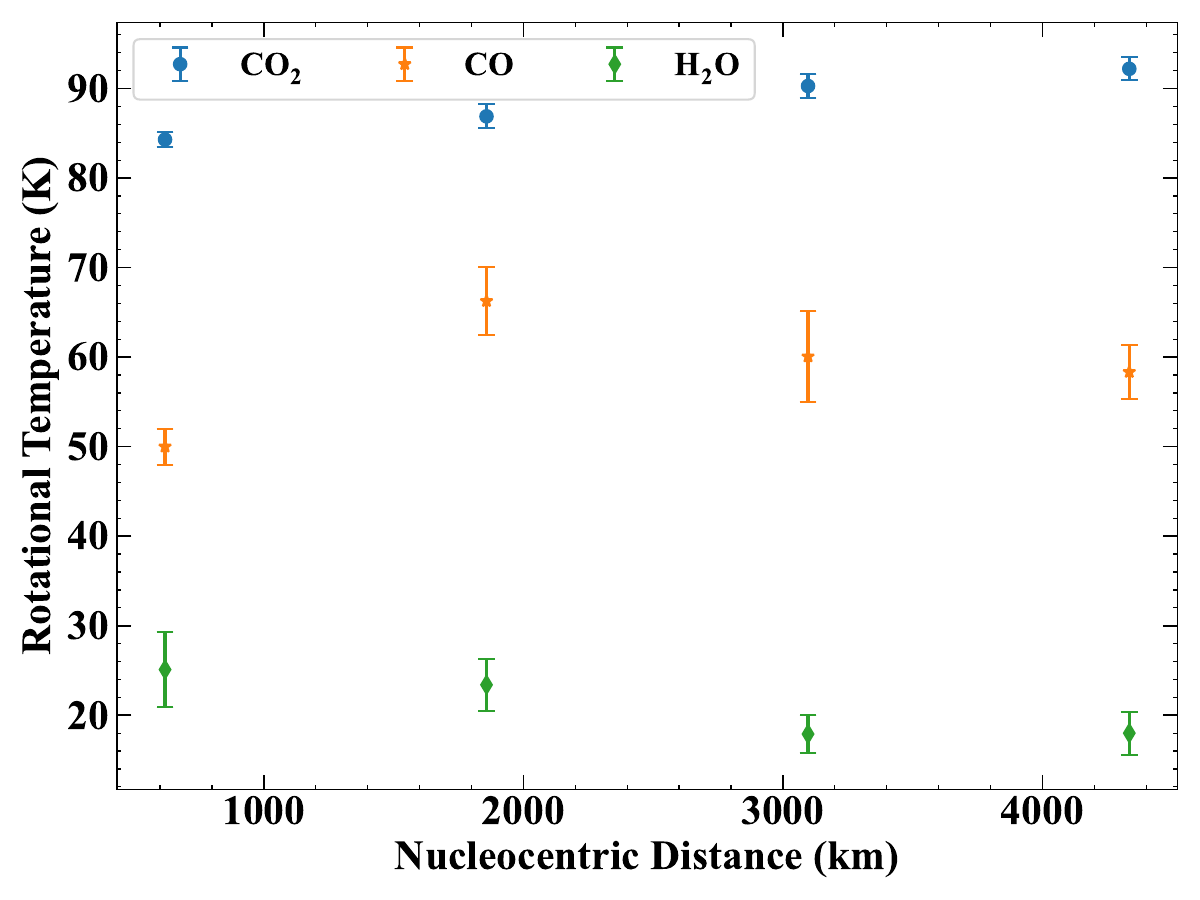}
\end{center}
\caption{Best fitting rotational temperatures for CO$_2$, CO, and H$_2$O as a function of sky-projected distance from the nucleus.}
\label{fig:tcurves}
\end{figure*}

For the $2.7$~$\mu$m region, H$_2$O spectral models were generated using PSG, while the continuum was determined using Equation \ref{eq:sigmoid}, combined with the $\lambda/\Delta\lambda$=5000 Solar spectral model of \citet{bohlin14}. Additional (weak) spectral contributions from coma CO$_2$ to this region were added based on our best fit to the 4.3~$\mu$m band.  The H$_2$O production rate and rotational temperature were retrieved using a Gaussian processes approach within the George software package \citep{ambikasaran15}, and uncertainties were derived using a Markov Chain Monte Carlo algorithm \citep[emcee;][]{foreman-mackey13}.  The retrieved rotational temperatures as a function of nucleocentric distance for each observed species are shown in Figure \ref{fig:tcurves}.

Examination of Figure~\ref{fig:spectrum} indicates the presence of $^{13}$CO$_2$ alongside $^{12}$CO$_2$, affording the first opportunity to test the $^{12}$CO$_2$/$^{13}$CO$_2$ ratio in an interstellar object. Compared to the much stronger $^{12}$CO$_2$ emission, the $^{13}$CO$_2$ band is only clearly detected in the the $0\farcs625$ nucleus-centered aperture. Unfortunately, the $^{12}$CO$_2$ in this region suffers from optical depth effects, and the low $(\lambda/\Delta\lambda\sim200)$ spectral resolution introduces additional difficulties in disentangling the contributions from each isotopologue. We therefore focused on analyzing the spectra from the first annular sector, where optical depth effects are reduced. We used the PSG to retrieve production rates $Q({\rm ^{12}CO_2})=(1.15\pm0.01)\times10^{27}$~s$^{-1}$ and $Q({\rm ^{13}CO_2})<1.50\times10^{25}$~s$^{-1}$. This corresponds to a $^{12}$C/$^{13}$C lower limit of $>$ 63 (3$\sigma$), which is formally consistent with the terrestrial value of 89. Additional observations of $^{13}$CO$_2$ in 3I/ATLAS at higher spectral resolution and signal-to-noise will be invaluable in separating it from its optically thick $^{12}$CO$_2$ counterpart, in order to place improved constraints on the CO$_2$ isotopic ratio.

\section{Active Sublimating Surface Areas}

We used the cometary ice thermal sublimation model of \citet{Cowan1979} to calculate the CO$_2$, H$_2$O, and CO active surface areas, assuming an infrared emissivity of 0.95, albedo of 5\% and nucleus radius of $<2.8$~km \citep{Jewitt2025}.  The Small-Bodies-Node/ice-sublimation code \citep{VanSelous2021} was used to calculate the average sublimation rate per unit area, $Z$, at $r_{\rm H}$ = 3.32 au, assuming a non-rotating, spherical nucleus. The active area is derived by dividing our terminal $Q$ values by $Z$, resulting in values of 3.1~km$^{2}$ for CO$_2$, 2.2~km$^2$ for H$_2$O, and 0.2~km$^2$ for CO. The active fractional area of the nucleus for each species is found by dividing these active areas by the (assumed) nucleus surface area. The corresponding lower limits on the active fractional area for each ice are $>3.1\%$, $>2.2\%$, and $>0.2\%$, respectively.

\end{document}